\documentclass[12pt]{article}
\usepackage{subcaption}
\usepackage{amssymb,amsmath}
\usepackage{graphicx}
\usepackage{float}
\usepackage{color}
\usepackage{array,multirow}
\usepackage[colorlinks=true
,urlcolor=blue
,citecolor=blue
,linkcolor=blue
,pagecolor=blue
,linktocpage=true
,pdfproducer=medialab
]{hyperref}
\usepackage[a4paper]{geometry}
\usepackage{cite}
\usepackage{placeins}
\usepackage[inline]{enumitem}
\usepackage{cancel}
\usepackage{arydshln}
\makeatletter \renewcommand{\@dotsep}{10000} \makeatother
\usepackage{indentfirst}
\def\beq{\begin{equation}}
\def\eeq{\end{equation}}
\usepackage[nodisplayskipstretch]{setspace}
\begin{document}
\begin{titlepage}
\pagestyle{empty}
\vspace*{0.2in}

\begin{center}
{\Large \bf Higgs boson in a flavor-extension of the CMSSM}\\
\vspace{1cm}
{\bf Surabhi Gupta$^{a}$\footnote{E-mail:sgupta2@myamu.ac.in }, 
Sudhir Kumar Gupta$^{a}$\footnote{E-mail:sudhir.ph@amu.ac.in},
and Keven Ren $^{b}$\footnote{E-mail:Keven.Ren@monash.edu}}
\vspace{0.5cm}
\begin{flushleft}
{\it $^a$Department of Physics, Aligarh Muslim University, 
Aligarh, UP--202002, India} \\
{\it $^b$School of Physics and Astronomy, Monash University, 
Melbourne, Victoria 3800, Australia}\\
\end{flushleft}
\end{center}

\vspace{0.5cm}
\begin{abstract}
Flavour-violating couplings of Higgs boson with stop and scharm quarks could be very important as in addition to lifting the mass of the Higgs boson by a few GeV, it could also play a vital phenomenological role in reducing the Supersymmetry breaking scale significantly. In this 
work, we investigate effects of such flavour-violating couplings within the Constrained Minimal Supersymmetric Standard Model (CMSSM) framework in the context of LEP data, Higgs data at the LHC, precision observables and the relic density of the dark matter using Bayesian statistics. Our 
detailed analysis reveals that the most probable values of $m_{0}$, $m_{1/2}$, $A_{0}$, $\tan{\beta}$, $\delta^{LR}_{ct}$ are expected to be around 4.83 TeV, 2.54 TeV, 1.90 TeV, 41.5, and 6.1$\times10^{-2}$, respectively, with flat priors. The corresponding values translate into 3.25 TeV, 2.13 TeV, 1.90 TeV, 44.7, and 5.9$\times10^{-2}$, respectively, if the natural priors are used. Furthermore, a comparison of our model with the CMSSM of flavour-conservation as the base model yields a Bayes factor of about 6 while taking into account all the experimental constraints used in our study. Our analysis also reflects that the lightest neutralino would have a mass of about 1 TeV.
\end{abstract}
\end{titlepage}

\section{Introduction}
Within the Standard Model (SM)~\cite{Djouadi:2005gi}, the Higgs mechanism is the underlying mechanism allowing for the electroweak (EW) symmetry breaking and thus generation of masses for the elementary particles. The resulting postulate of a scalar fundamental particle is one of the major reasons for construction of the Large Hadron Collider (LHC). In 2012, the ATLAS and CMS collaborations have recorded the 
observation of a spin-zero particle with qualities consistent with that of the Higgs boson~\cite{Aad:2015zhl}. Under the Constrained Minimal Supersymmetric Standard Model 
(CMSSM)~\cite{Kane:1993td,allenach,Balazs:2013qva,Fowlie:2012im,Athron:2017fxj,GAMBIT:2017snp,Ellis:2018jyl} 
framework, only five parameters are required to generate the 
supersymmetric spectrum. However, it remains difficult for CMSSM to justify a Higgs mass of 125 GeV with reasonable assumptions. Recent experimental results for the non-observance of a supersymmetric particle below 1 TeV bring additional stress to the theory as the little hierarchy issue becomes more prominent with heavier stops. The amount of 
fine-tuning required increases with the energy gap between the Supersymmetry (SUSY)~\cite{Martin:1997ns, Tata:1997uf, Drees:1996ca, Aitchison:2005cf, Fayet:2015sra, Djouadi:2005, Cane:2019ac, Allanchach:2019wrx,Chung:2003fi} breaking scale and the EW regime. Therefore, it is desirable to search beyond the SM phenomenon which may propose alternative solutions to accommodate a Higgs mass of 125 GeV without stretching this gap. Models such as Peccei-Quinn extension to the SM~\cite{Balazs:2012bx} and non-minimal flavour violation 
(NMFV)~\cite{Heinemeyer:2004by,Bozzi:2007me,AranaCatania:2011ak,AranaCatania:2012sn,Arana-Catania:2014ooa,Kowalska:2014opa,DeCausmaecker:2015yca,Bernigaud:2018qky,Hu:2019heu, 
Hahn:2005qi, Carena:2006ai, Bruhnke:2010rh, Herrmann:2011xe, 
Fuks:2011dg, Fuks:2008ab, delAguila:2008iz, Bernigaud:2018vmh,Gupta:2022bjx} are examples of methods one may utilise to explain a 125 GeV Higgs mass. The first work on NMFV has been presented in Ref.~\cite{Heinemeyer:2004by}, where the authors have studied the effects of NMFV in the MSSM involving up-type squarks of the second and third generation in the left-left (LL) sector and have shown contributions to 
the electroweak observables and the lightest MSSM Higgs boson mass. Another work has been published in Ref.~\cite{Bozzi:2007me} about the contributions
of the NMFV SUSY in context of the hadroproduction and 
decay of sparticles. In a work by Arana-Catania et al.~\cite{AranaCatania:2011ak,AranaCatania:2012sn,Arana-Catania:2014ooa}, the authors have contended that flavour-changing couplings, particularly within the scharm-stop sector, 
have the capacity to increase the Higgs mass by up to 10 GeV. 
The additional flavour-violating interactions are also expected to reduce the unified masses, therefore bringing the squark masses closer to the EW scale. The connection has been beheld between the flavour-violating terms of the squark and the slepton sectors at the Grand Unified Theory (GUT) 
and the TeV scale in the NMFV SUSY framework~\cite{Bernigaud:2018qky}.
Additional works in NMFV have been implemented in other directions including 
Higgs boson decays~\cite{Hahn:2005qi}, Gauge-mediated SUSY Breaking (GMSB) model~\cite{Fuks:2008ab}, Anomaly-mediated SUSY Breaking (AMSB) model~\cite{Fuks:2011dg}, $Z_3$ invariant Next-to MSSM (NMSSM) scenario~\cite{Hu:2019heu}, and CMSSM model~\cite{Gupta:2022bjx}. Flavour-changing neutral-current process~\cite{Chung:2003fi} similar to the NMFV context has been explored in the hybrid gauge-gravity 
model~\cite{Hiller:2008sv} and the top-quark processes at the 
LHC~\cite{Guasch:1999jp, Cao:2006xb, Cao:2007dk}. Significant contribution of 
the supersymmetric particles towards the decays of flavour-changing neutral 
MSSM Higgs bosons to the second and third generation quarks has also been 
studied in Refs.~\cite{Curiel:2002pf,Curiel:2003uk,Bejar:2004rz}. Motivated 
by the premise of Arana-Catania et al., we investigate a CMSSM scenario 
extended with non-minimal flavour-violating interaction and 
then compare it with the vanilla CMSSM case. Both models are subjected to 
various experimental constraints and good fitting regions of the 
parameter space are found and compared. A Bayesian analysis is then 
provided to give a quantitative basis for the plausibility of the models 
against each other. \\ The rest of the paper is organised as follows. 
Section II discusses the NMFV extension to CMSSM. Section III includes 
our method, Bayes statistics and the constraints applied. Section IV 
gives the detailed results and Bayesian analysis. Section V is the 
conclusion.

\section{Flavour--violation in MSSM and the Higgs boson} 
Within MSSM the flavour-violating terms could typically arise due to 
the RG flow of various squark masses from the SUSY-breaking scale 
down to the EW scale. However such terms are heavily suppressed. Under the 
NMFV scheme, the off-diagonal flavour-violating terms of the squark mass 
matrices under the super-CKM basis are not suppressed at lower energy scales,
particularly in the top-charm sector~\cite{Atwood:2013ica}. Therefore, 
considering the flavour violation in the top-charm sector, which could 
possibly give rise to a few extra GeV contribution to the Higgs boson mass 
and thus reduce the SUSY-breaking scale significantly. The mass matrices
of the up-type squarks could be described as

\begin{eqnarray} 
\label{eq:y:1}
{\cal M}^2_{~\tilde{U}_X}  	=	\begin{pmatrix}
	M^{2}_{\tilde{U}_{X11}}	& 0 										& 0 										\\
	0 					& M^{2}_{\tilde{U}_{X22}} 						& \delta^{XX}_{23} M_{\tilde{U}_{X22}}M_{\tilde{U}_{X33}}	\\
	0 					& \delta^{XX}_{23} M_{\tilde{U}_{X22}}M_{\tilde{U}_{X33}} 	& M^{2}_{\tilde{U}_{X33}}. 						\\
						\end{pmatrix},
\end{eqnarray}
with the up-type trilinear coupling in terms of flavour-violating 
off-diagonal terms could be given as

\begin{eqnarray} 
\label{eq:y:2}
v_{1}{\cal A}^{u}		=	\begin{pmatrix}
	0					& 0										& 0										\\
	0					& 0										& \delta^{LR}_{ct} M_{\tilde{U}_{L22}}M_{\tilde{U}_{R33}}	\\
	0					& \delta^{RL}_{ct} M_{\tilde{U}_{R22}}M_{\tilde{U}_{L33}} 	& m_{t}{A}_{t}
				\end{pmatrix}.
\end{eqnarray}
\noindent
In Eq.~\eqref{eq:y:1}, the dimensionless coefficient $\delta^{XY}_{ij}$
is the flavour-violating coupling where $\emph{X}$ is either chiral 
$\emph{L}$ (left) or $\emph{R}$ (right), $i$, $j$ represent the generations 
of the squarks, and $M_{\tilde{U}_{Xij}}$ is the up-type squark mass. In 
Eq.~\eqref{eq:y:2}, $v_1$ is the vacuum expectation value (VEV) of the $H_u$,
matrix ${\cal A}^u$ includes values of the Yukawa-type couplings of 
the Higgs and the up-type squarks, $m_t$ denotes the top-quark mass, 
and $A_t$ represents the trilinear coupling of the stop. 
The corresponding down-type mass matrices ${\cal M}^2_{\tilde{D}_X}$ 
and down-type trilinear coupling $v_{2}{\cal A}^{d}$, where $v_2$ 
represents the VEV of the $H_d$ and matrix ${\cal A}^d$ includes values 
of the Yukawa-type couplings of the Higgs and the down-type squarks,
completely analogous to Eqs.~\eqref{eq:y:1} and \eqref{eq:y:2}, 
respectively and could be found by replacing the respective indices. 
The contributions towards the Higgs masses are expected to occur in 
the $\emph{LR}$/$\emph{RL}$ category, as the flavour-violating coupling
is now being factored directly into the Yukawa-type coupling involving 
the Higgs boson and squarks~\cite{AranaCatania:2011ak,AranaCatania:2012sn,Arana-Catania:2014ooa}. The suppression of flavour-violating interaction involving the 
first generation squarks is due to existing experimental data already 
restricting its possible range. In this investigation, only the NMFV 
effects in the second and third generation squarks in the $\emph{LR}$ 
sector are considered due to the most significant contribution provided 
to the Higgs mass. As a result the SUSY breaking scale is significantly 
reduced relative to the $\emph{RL}$, $\emph{LL}$, and $\emph{RR}$ sectors. 
In~\cite{Atwood:2013ica}, the extent of FV within the top-charm sector 
has been constrained to $\left|{\delta}^{XY}_{ct}\right| \lesssim 0.5$ at 
a 1$\sigma$ confidence level using the measurements on the decay width 
of top-quark, $\Gamma_t = 1.42^{+0.19}_{-0.15}$ GeV~\cite{Zyla:2020zbs}. 
In the context of the CMSSM the flavour-violating coupling appears in 
the one-loop correction to the Higgs mass, which can be expressed as
\begin{eqnarray}
\label{eq:y:3}
\triangle m_{h}\left(\delta^{LR}_{ct}\right) \equiv m^{NMFV}_{h}\left(\delta^{LR}_{ct}\right)-m^{CMSSM}_{h},
\end{eqnarray}
where $\triangle m_{h}\left(\delta^{LR}_{ct}\right)$ is the correction to the Higgs mass due to the inclusion of flavour-violating coupling.  If $\delta^{LR}_{ct}$ = 0 then this correction vanishes: $m^{NMFV}_{h}\left(\delta^{LR}_{ct}\right)$ = $m^{CMSSM}_{h}$. In this work we use {\tt FeynHiggs}~\cite{Bahl:2018qog,Bahl:2017aev, Bahl:2016brp, Hahn:2013ria, Frank:2006yh, Degrassi:2002fi, Heinemeyer:1998np, Heinemeyer:1998yj} for the numerical computation of the parameters $m^{NMFV}_{h}\left(\delta^{LR}_{ct}\right)$ and $m^{CMSSM}_{h}$. Further details on Eq.~\eqref{eq:y:3} can be found in Refs.~\cite{AranaCatania:2011ak,AranaCatania:2012sn,Arana-Catania:2014ooa}.

\section{\label{numerical}Numerical analysis}

We use a Bayesian analysis to quantitatively evaluate the plausibility 
of our model. Comparing theoretical predictions of our model to 
experimental observations, we can determine regions 
in the parameter space where our model best agrees with the 
experiment. To this end, we introduce the likelihood function 
for a point $x = \{x_1, x_2, ..., x_n\}$ in the parameter space as
	\begin{eqnarray}
	\label{eq:y:4}
		{\cal L}_{i}\left(x\right)=\dfrac{1}{\sqrt{2\pi}\sigma_{i}}\exp\left(-\dfrac{\chi^{2}_{i}}{2}\right).
	\end{eqnarray}
\noindent
Here, for each observables ${\cal O}_{i}$, $\chi_{i}^{2}$ is defined as
	\begin{eqnarray}
	\label{eq:y:5}
	\chi_{i}^{2}=\dfrac{\left({\cal O}_{i}\left(x\right)-{\cal O}_{i}^{exp}\right)^{2}}{\sigma_{i}^{2}},
	\end{eqnarray}
\noindent	
where ${\cal O}_{i}^{exp}$ is the experimentally measured value of the observable, 
${\cal O}_{i}(x)$ is the corresponding theoretical prediction calculated at a point in the 
parameter space, and $\sigma_{i}$ is the corresponding uncertainty. The 
likelihood quantifies the agreement of the theoretical 
prediction with experiment. To include all observables considered in 
our analysis, as shown in Table~\ref{tab:table3}, we formed the composite likelihood 
as the product of all ${\cal L}_{i}$,
	\begin{eqnarray}
	\label{eq:y:6}
		{\cal L} = \displaystyle\prod_{i} {\cal L}_i.
	\end{eqnarray}
\noindent
The posterior probability distribution of a given parameter $x_j \in x$ is calculated 
by marginalising over the rest of the parameter space,
\begin{eqnarray}
\label{eq:y:7}
 \mathcal P(x_j) = \frac{\int {\cal L}\xi \prod_{i\not=j} dx_i}{\mathcal {E}} ,
\end{eqnarray}
where $\xi$ denotes our selection of priors, and the evidence is defined as
\begin{eqnarray}
\label{eq:y:8}
{\mathcal {E}} = {\int{\cal L} \xi dx} .
\end{eqnarray}
The evidence is essential for calculating the 
Bayes factor, ${\it Q_{Bayes}}$, which is a quantitative measure 
of the plausibility of the NMFV CMSSM compared to CMSSM without NMFV couplings. 
The Bayes factor is written as
\begin{eqnarray}
\label{eq:y:9}
 Q_{Bayes} = \log_{10} \left(\frac{{\mathcal {E}}_{NMFV}}{{\mathcal {E}}_{CMSSM}}\right).
\end{eqnarray}

			\begin{table}
			\begin{center}
			\begin{tabular}{ llr}							
			{\bf Particle\quad}		& {\bf Bound}		& {\bf Source}		\\\hline\hline
            $m_{t}$				& $172.89\pm0.59$ GeV	&\cite{Zyla:2020zbs}     \\
			$m_{h}$				& $ > 114.4 $ GeV	& \cite{Schael:2006cr}	\\
			$m_{\tilde{\chi}^{0}_{1}}$	& $ > \frac{m_{Z}}{2}$	& \cite{Zyla:2020zbs}	\\
			$m_{\tilde{\chi}^{\pm}_{1}}$	& $ > 103.5 $ GeV	& \cite{Zyla:2020zbs}\\
		\hline\hline
			\end{tabular}
	\caption{LEP bounds on Higgs and sparticles masses.}
	\label{tab:table1}
	   \end{center}
		\end{table}

We analyse the NMFV parameter space subjected to various constraints 
including LEP data on Higgs mass and sparticles masses, Higgs observables,
precision observables, and the abundance of dark matter at a 
2.5$\sigma$ confidence level. Details of these constraints and their 
experimental values can be found in Tables~\ref{tab:table1} and \ref{tab:table2}. 
The experimental data account for the Higgs observables~\cite{Aad:2015zhl, Zyla:2020zbs} obtained from the ATLAS and CMS collaborations. The ${\mathcal R}$ is 
the ratio of overall rate process of SUSY over SM counterpart.

The decay rate 
	\begin{eqnarray}
		{\mathcal R}_{gg\gamma\gamma} =	\left[ \dfrac{\sigma_{pp \rightarrow h}\times \Gamma^{h\rightarrow\gamma\gamma}}{\Gamma_{h}} \right]_{\frac{SUSY}{SM}} 
	\end{eqnarray}
is enhanced in supersymmetry, generating more $gg \rightarrow \gamma\gamma$ events than predicted by SM. We assume $\sigma_{pp \rightarrow h}$ $\simeq$ $\Gamma^{h\rightarrow gg}$ due to dominant Higgs production at the LHC by the gluon fusion.  Above $\Gamma^{h\rightarrow\gamma\gamma}$ is the partial decay width, and $\Gamma_{h}$ is the total decay width of Higgs boson. Similarly, the ${\mathcal R}$ value can be evaluated for NMFV over its SM counterpart for the processes $gg \rightarrow \gamma\gamma$, $gg \rightarrow 2l2\nu$, $gg \rightarrow 4l$, $gg \rightarrow bb$ and $gg\rightarrow
\tau\tau$.
       		\begin{table}
       		 \begin{center}
			\begin{tabular}{llll}							
{\bf Constraint\quad}	&			{\bf Observable\quad}		& {\bf Quantity}	&	{\bf Source}		\\\hline\hline
Higgs observables (HO)			&$m_{h}$			& $ 125.09 \pm 0.24 $ GeV	&\cite{Aad:2015zhl}		\\
			&${\mathcal R}_{gg\gamma\gamma}$ & $1.11\pm 0.10$ 		    &\cite{Zyla:2020zbs}		\\
			&${\mathcal R}_{gg2l2\nu} $		& $1.19\pm 0.12 $		&\cite{Zyla:2020zbs}		\\
			&${\mathcal R}_{gg4l}$			& $1.20 \pm 0.12 $			&\cite{Zyla:2020zbs}	\\
	        &${\mathcal R}_{ggbb}$			& $1.04 \pm 0.13 $			&\cite{Zyla:2020zbs}	\\			
		    &${\mathcal R}_{gg\tau\tau}$			& $1.15 \pm 0.16$   &\cite{Zyla:2020zbs}	\\
\hline
\hline
Precision observables (PO)		&	$\triangle \rho$				& $ 0.00038 \pm 0.0002$				& \cite{Zyla:2020zbs}	\\
		&	$\triangle a_{\mu}$					& $ (2.51\pm 0.59) \times 10^{-9} $		& \cite{Abi:2021gix}	\\
		&	$Br(b\rightarrow s\gamma)$	& $ (3.32\pm 0.15) \times 10^{-4}$		& \cite{Zyla:2020zbs,Amhis:2019ckw}	\\
		\hline\hline
Dark matter (DM) & $ \Omega_{\chi}h^{2}$ & $0.1197\pm0.0022$& \cite{Planck:2015fie}\\						\hline\hline
	\end{tabular}
	\caption{List of experimental observables applied in our analysis.}
	\label{tab:table2}
	    \end{center}
		\end{table}

\section{Results and Discussions}
In this section, we present our results in the form of the posterior 
probability distributions under various experimental constraints. The 
models examined in this discussion are cases of CMSSM with NMFV in 
the $\emph{LR}$ sector of the scharm-stop flavour-violating interaction 
and CMSSM with no NMFV interactions. We examine our study only in the 
scharm-stop sector as the contributions of couplings involving up-quark 
with top-quark or charm-quark are suppressed. Moreover, the coupling 
$\delta^{LR}_{ct}$ provides the most significant contribution 
to the Higgs mass as compared to other couplings like 
$\delta^{RL}_{ct}$, $\delta^{LL}_{23}$, and $\delta^{RR}_{ct}$ as 
discussed in Refs.~\cite{AranaCatania:2011ak,AranaCatania:2012sn,Arana-Catania:2014ooa} 
and also reduces the SUSY breaking scale significantly. The ATLAS 
and the CMS collaborations attained the combined mass measurement of the 
lightest Higgs boson as 125.09 $\pm$ 0.24 GeV~\cite{Aad:2015zhl}. 
Without modifications to CMSSM, the theory experiences difficulty in
achieving this value without pushing the supersymmetric breaking scale 
to problematic ranges. We allude to this problem with the introduction 
of NMFV in the scharm-stop sector, where it can be seen to favourably 
impact the unified supersymmetric parameters. The 125 GeV Higgs mass can 
now be accomplished effortlessly following one-loop correction to the 
Higgs mass under the influence of NMFV coupling. In the calculation of 
the Bayes factor, the full scope of this study is included for brevity. 
Here we employ a random scan technique over the entire parameter space 
of CMSSM with NMFV as follows

\begin{itemize}
\item $m_0 \in [0.01,6] $ TeV,
\item $m_{1/2} \in [0.01,6] $ TeV,
\item $A_{0} \in [-6,6]$ TeV,
\item $tan\beta \in [0,60]$,
\item sgn$(\mu)$ = +1,
\item$\delta^{RL}_{ct}$  $\in$ [$-$0.07,0.07].
\end{itemize}
\noindent
The parameters sample includes the usual five parameters of CMSSM, 
namely $m_{0}$ unified scalar mass, $m_{1/2}$ unified gaugino mass, 
$A_{0}$ common trilinear coupling parameter, $\tan{\beta}$ ratio of VEVs 
of up- and down-type Higgs bosons, $\Big(\dfrac{v_1}{v_2}\Big)$, and 
$\operatorname{sgn}(\mu)$ sign associated with the Higgsino mass 
parameter $\mu$ along with a flavour-violating coupling 
$\delta^{LR}_{ct}$. The entirety of the supersymmetric spectrum is 
created using these terms and the results are filtered using various 
experimental constraints.

\noindent
Procuring the aforementioned parameters, we would use {\tt 
Softsusy}~\cite{Allanach:2001kg} to generate the spectrum of 
supersymmetric particles. An additional parameter involved in this step 
is defining the mass of the top quark. {\tt 
FeynHiggs}~\cite{Bahl:2018qog,Bahl:2017aev, Bahl:2016brp, Hahn:2013ria, 
Frank:2006yh, Degrassi:2002fi, Heinemeyer:1998np, Heinemeyer:1998yj} is 
in conjunction with {\tt Softsusy}~\cite{Allanach:2001kg} that is 
used in calculating the various Higgs observables such as masses,
mixings, and branching ratios and $\rho$-parameter up to the 
two-loop level. {\tt Superiso}~\cite{Arbey:2018msw} is used to 
calculate the relevant B-physics observables and the muon anomalous 
magnetic moment and, in addition, {\tt micromegas}~\cite{Belanger:2004yn, Belanger:2001fz} is employed to compute the relic density of the neutralino 
dark matter using the spectrum generator.

\begin{figure}[ht]
\begin{subfigure}{.54\textwidth}
  \centering
  \includegraphics[width=.99\linewidth,height=15em]{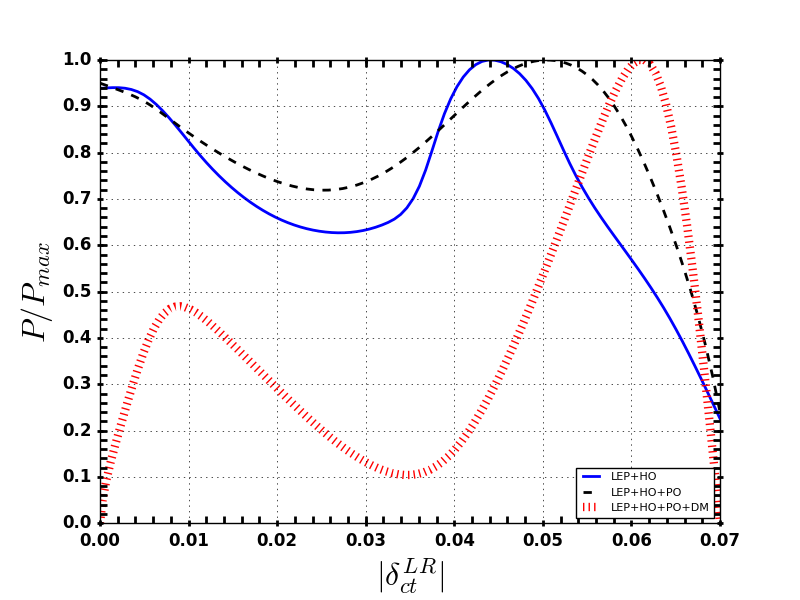}  
  \caption{}
  \label{fig:sub-first}
\end{subfigure}
\hspace{-2.4em}
\begin{subfigure}{.48\textwidth}
  \centering
  \includegraphics[width=.99\linewidth,height=15em]{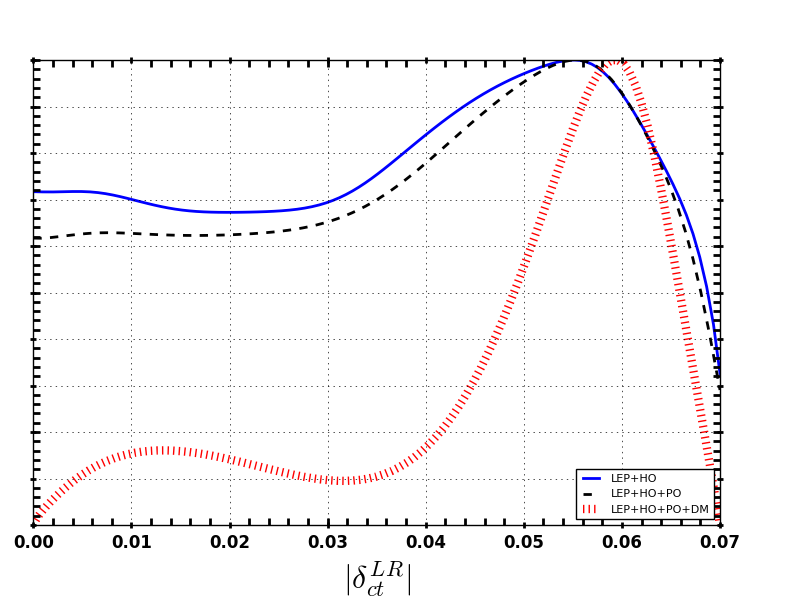}  
  \caption{}
  \label{fig:sub-second}
\end{subfigure}
\caption{Posterior probability distributions of $\delta^{LR}_{ct}$ flavour-violating coupling of NMFV for different sets of constraints applied with (a) flat and (b) natural priors. The blue solid line indicates LEP data and Higgs observables constraints, the black dashed line signifies the constraints of LEP data, Higgs observables, and precision observables and the red dotted line includes the constraints of LEP data, Higgs observables, precision observables, and the relic density of the dark matter. The description of constraints is presented in Tables~\ref{tab:table1} and \ref{tab:table2}.}
\label{fig:1}
\end{figure}

\begin{figure}
\includegraphics[angle=0,width=1.0\linewidth,height=24em]{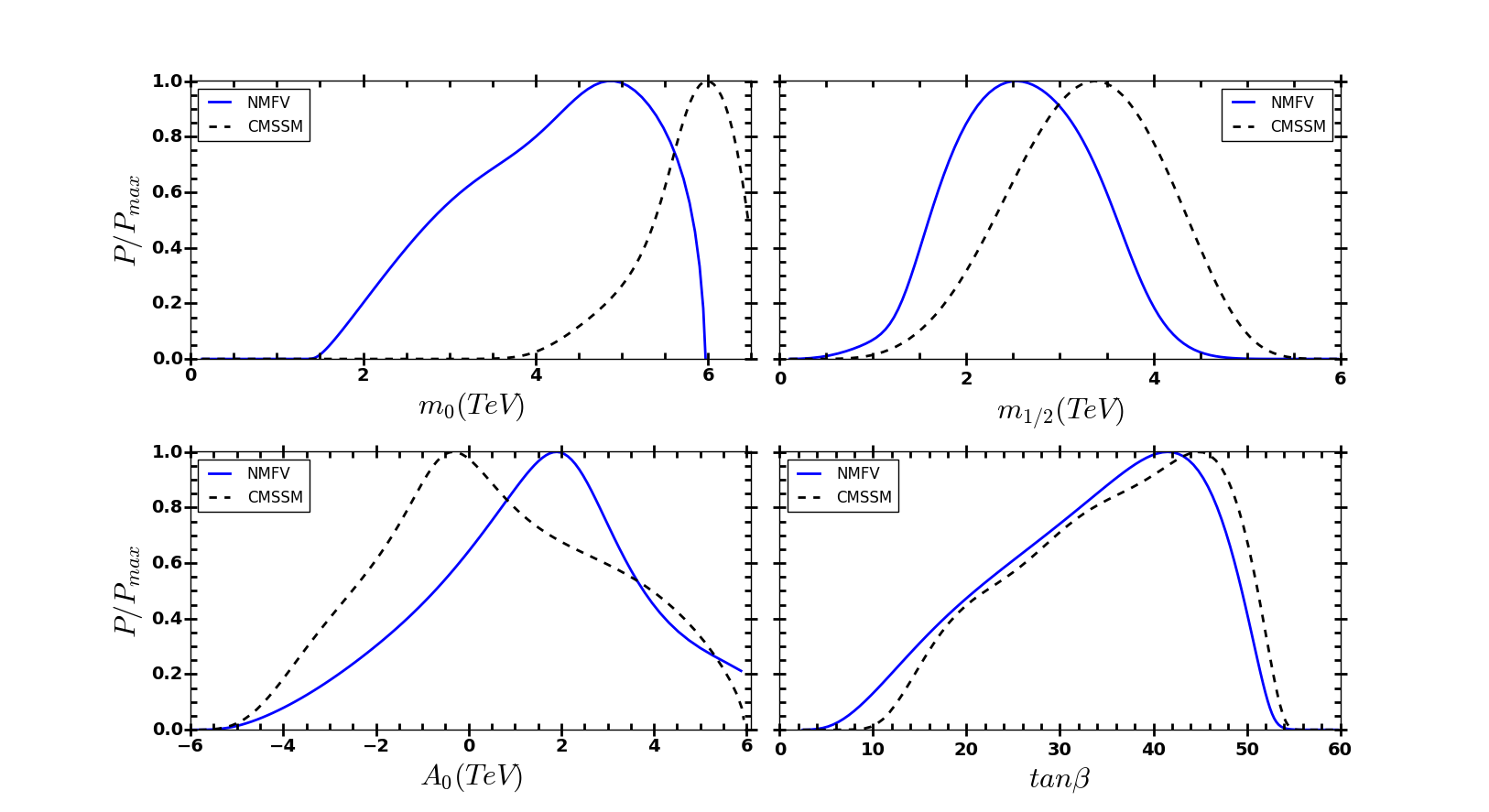}
\caption{Posterior probability distributions of CMSSM parameters taking into account the constraints of LEP data, Higgs observables, precision observables, and the relic density of the dark matter using flat priors. The blue solid line illustrates the NMFV while the black dashed line depicts the CMSSM.}
\label{fig:2}
\end{figure}

\begin{figure}
\includegraphics[angle=0,width=1.0\linewidth,height=24em]{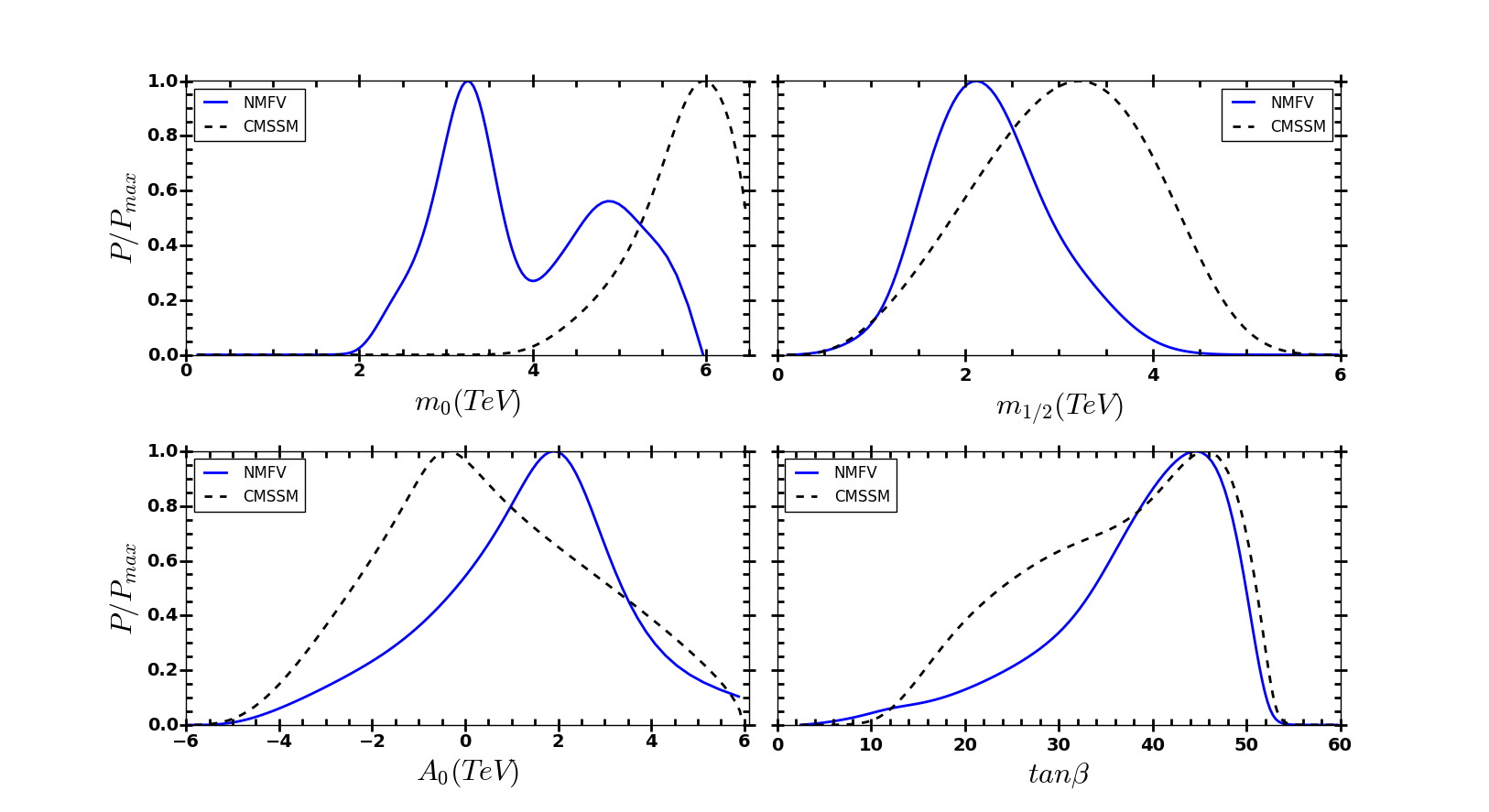}
\caption{Posterior probability distributions of CMSSM parameters involving the constraints of LEP data, Higgs observables, precision observables, and the relic density of the dark matter utilising natural priors. The colour convention follows the same as in Figure~\ref{fig:2}.}
\label{fig:3}
\end{figure}

\begin{figure}
\begin{centering}
\includegraphics[angle=0,width=0.32\linewidth,height=12em]{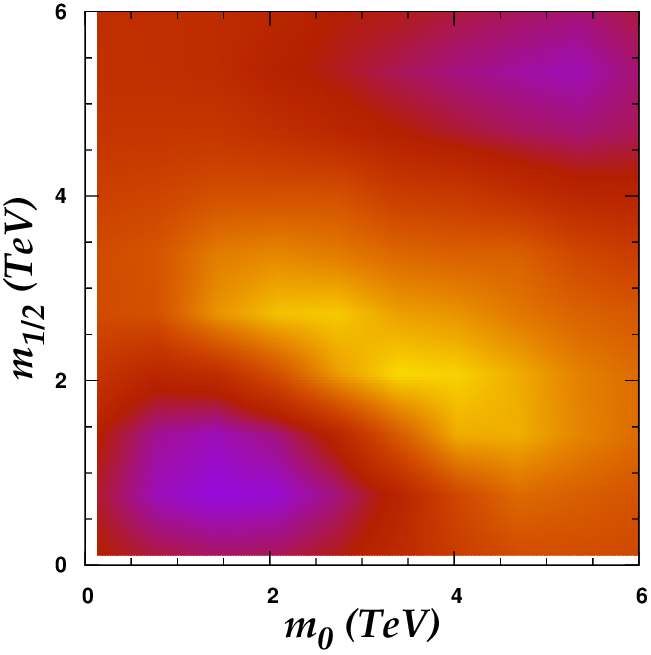}
\hspace{-0.7em}
\includegraphics[angle=0,width=0.32\linewidth,height=12em]{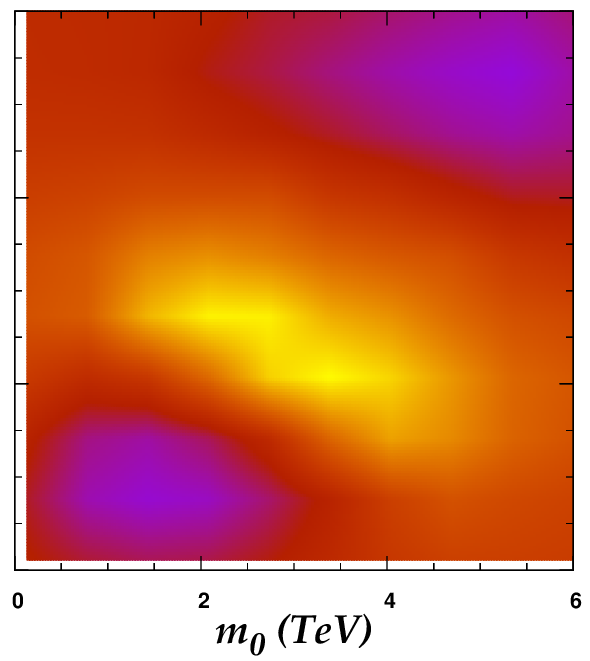}
\hspace{-0.8em}
\includegraphics[angle=0,width=0.32\linewidth,height=12em]{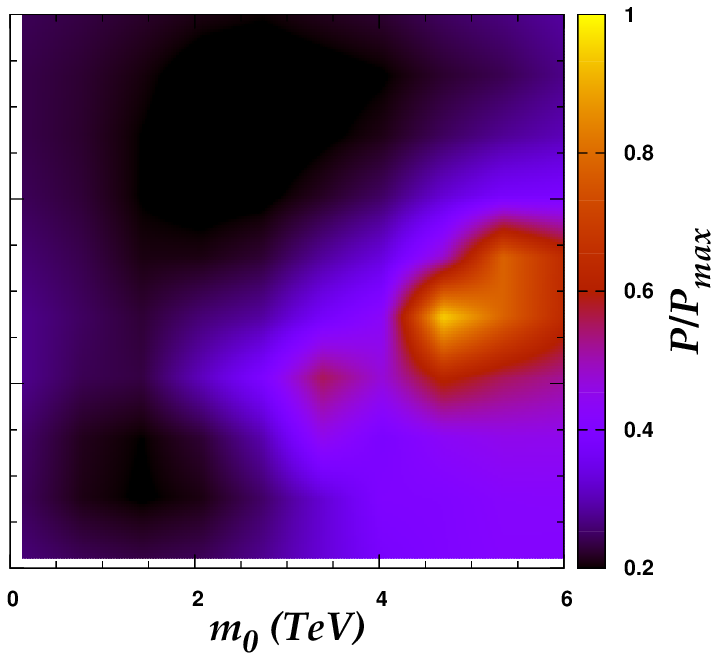}
\caption{Heatmaps of $m_0, m_{1/2}$ and posterior probability with flat priors for LEP$+$HO (left), LEP$+$HO$+$PO (middle), and LEP$+$HO$+$PO$+$DM (right) constraints in the $\emph{LR}$ sector of the scharm-stop flavour-violating interaction of the NMFV framework.}
\label{fig:4}
\end{centering} 
\end{figure}

\begin{figure}
\begin{centering}
\includegraphics[angle=0,width=0.323\linewidth,height=12.2em]{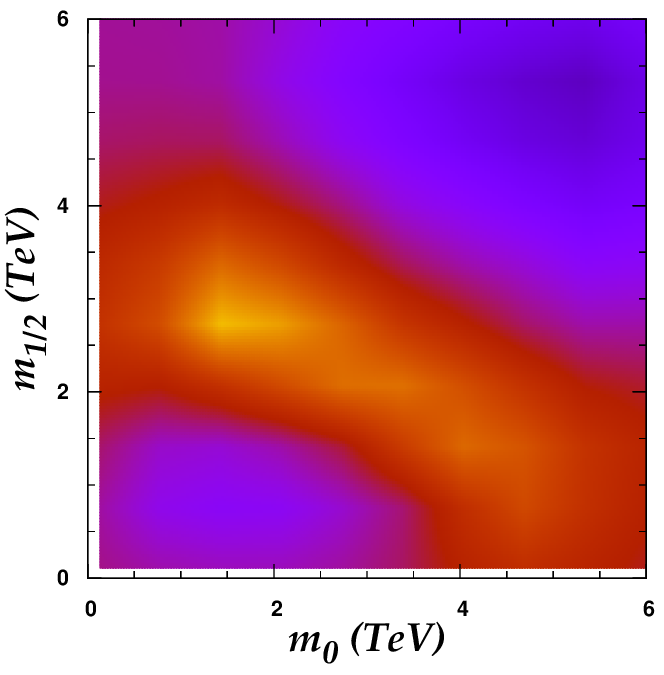}
\hspace{-0.8em}
\includegraphics[angle=0,width=0.30\linewidth,height=12em]{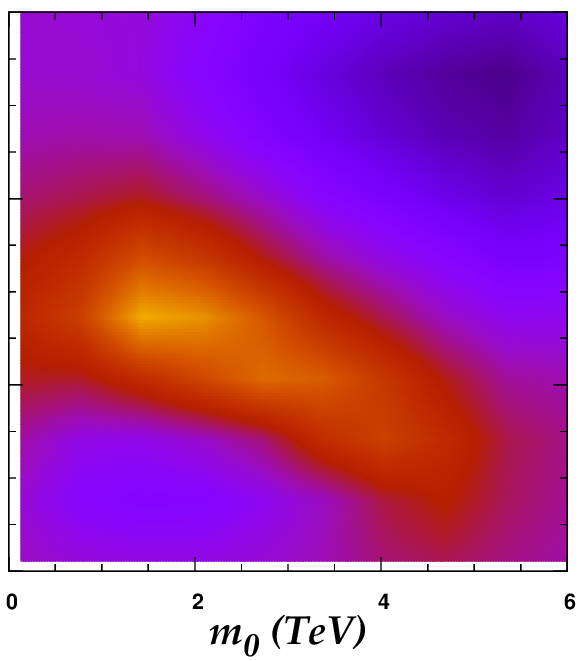}
\hspace{-0.7em}
\includegraphics[angle=0,width=0.32\linewidth,height=12em]{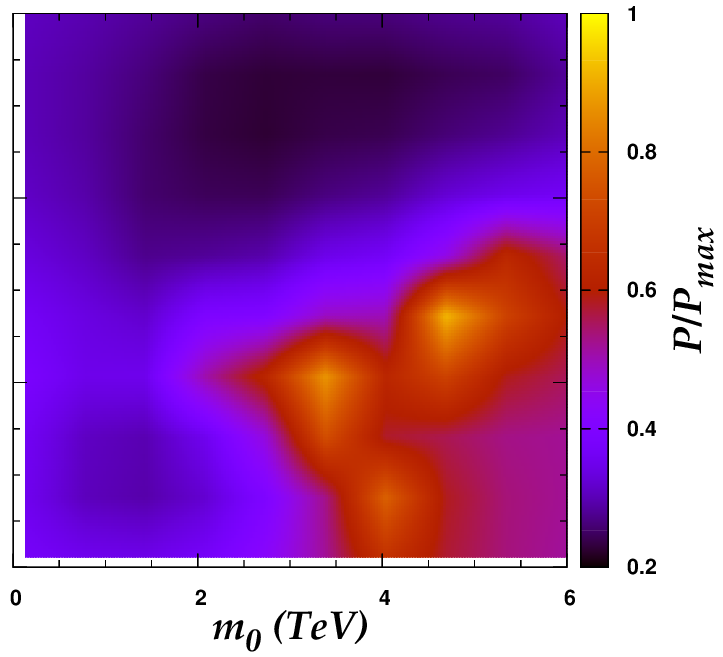}
\caption{Heatmaps of $m_0, m_{1/2}$ and posterior probability with natural priors for LEP$+$HO (left), LEP$+$HO$+$PO (middle), and LEP$+$HO$+$PO$+$DM (right) constraints in the $\emph{LR}$ sector of the scharm-stop flavour-violating interaction of the NMFV framework.}
\label{fig:5}
\end{centering} 
\end{figure}

\begin{figure}
\begin{centering}
\includegraphics[angle=0,width=0.32\linewidth,height=10em]{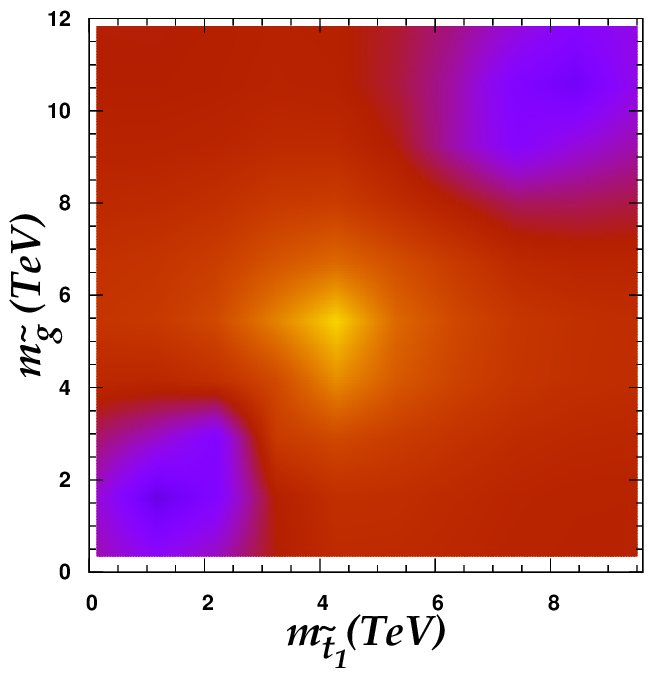}
\hspace{-0.5em}
\includegraphics[angle=0,width=0.32\linewidth,height=10em]{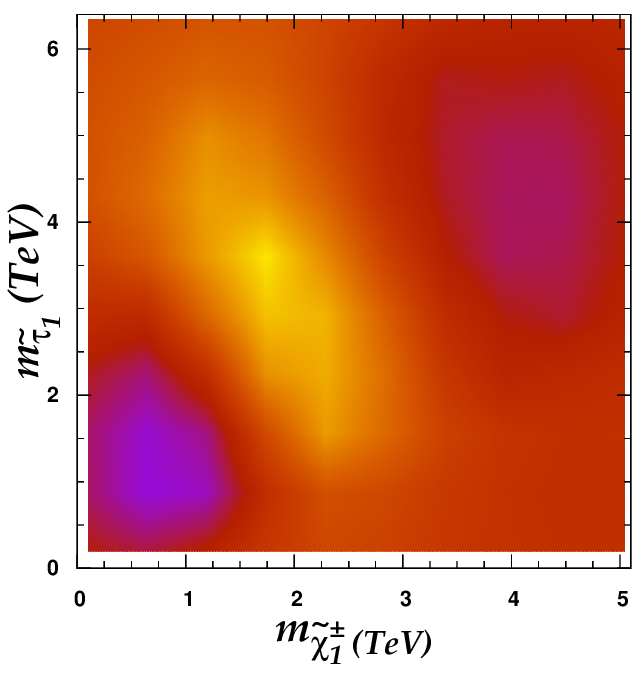}
\hspace{-0.5em}
\includegraphics[angle=0,width=0.33\linewidth,height=10em]{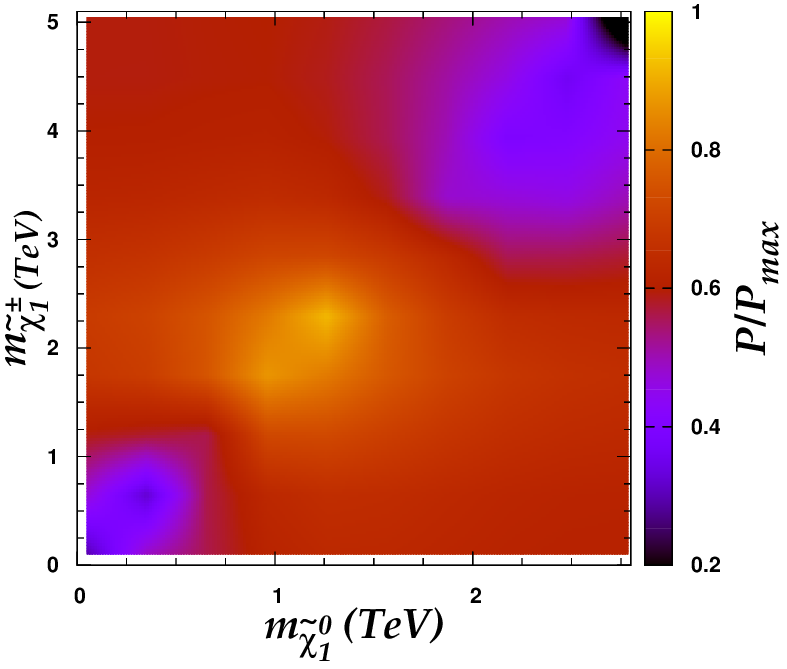}
\includegraphics[angle=0,width=0.32\linewidth,height=10em]{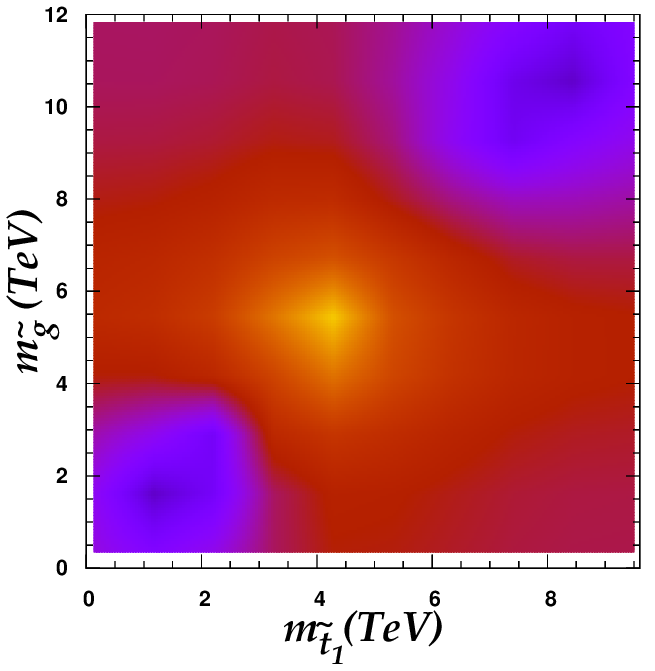}
\hspace{-0.5em}
\includegraphics[angle=0,width=0.32\linewidth,height=10em]{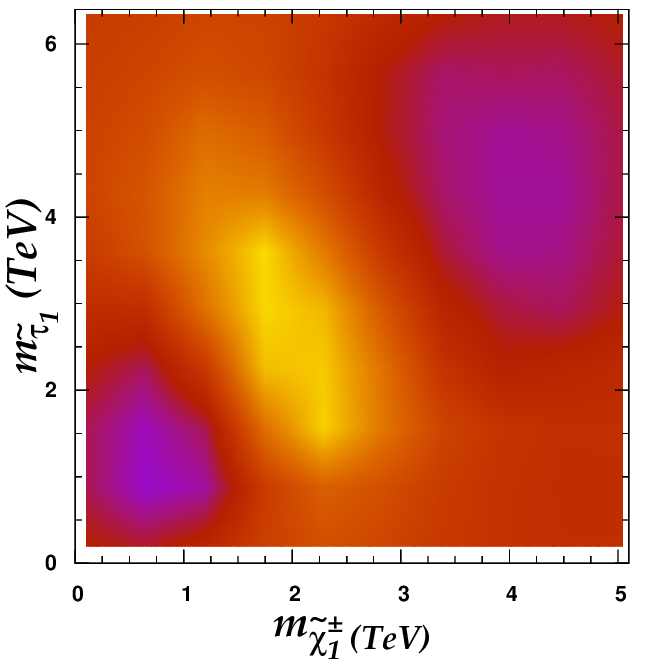}
\hspace{-0.5em}
\includegraphics[angle=0,width=0.33\linewidth,height=10em]{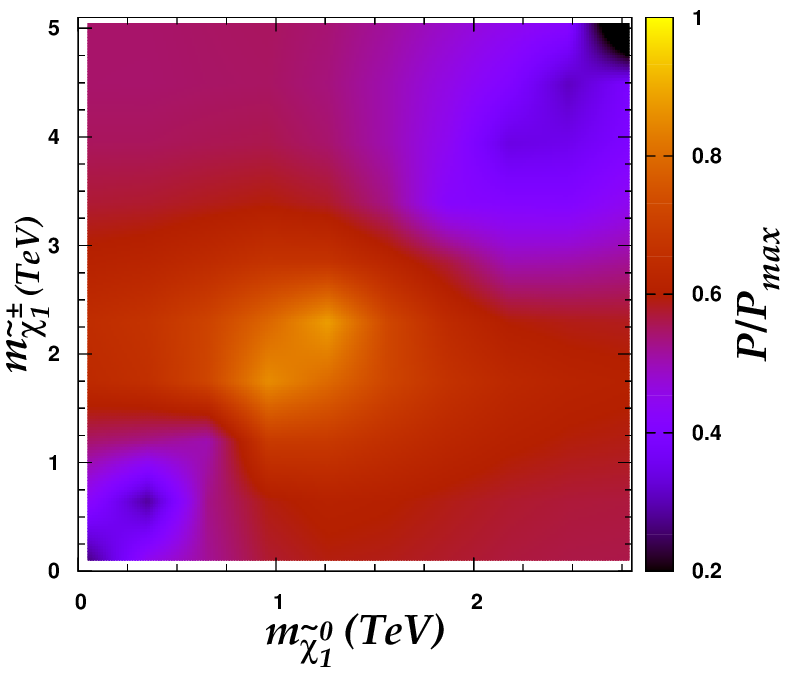}
\includegraphics[angle=0,width=0.32\linewidth,height=10em]{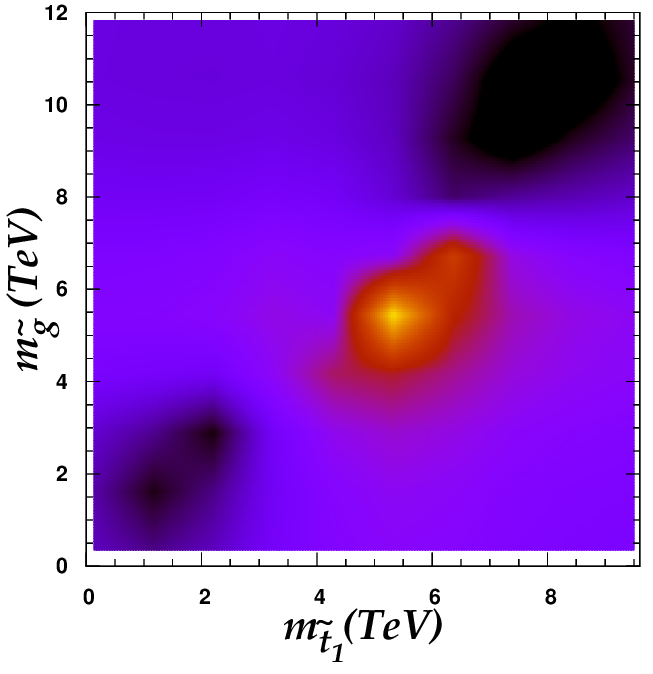}
\hspace{-0.5em}
\includegraphics[angle=0,width=0.32\linewidth,height=10em]{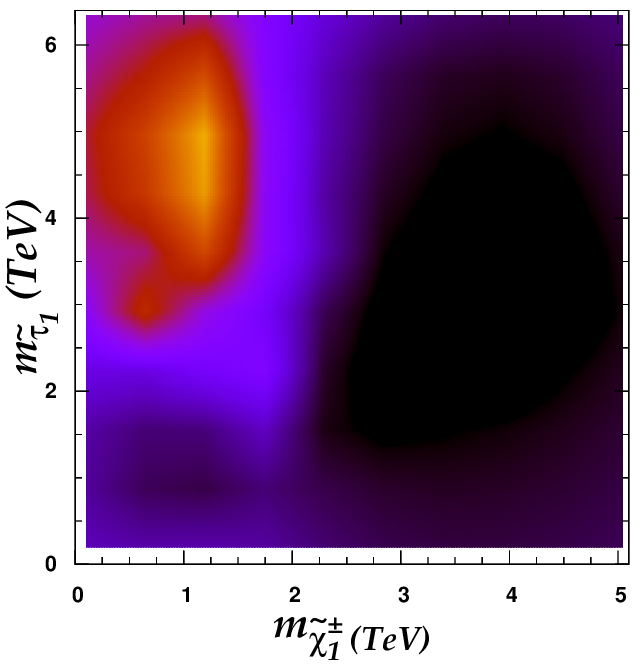}
\hspace{-0.5em}
\includegraphics[angle=0,width=0.33\linewidth,height=10em]{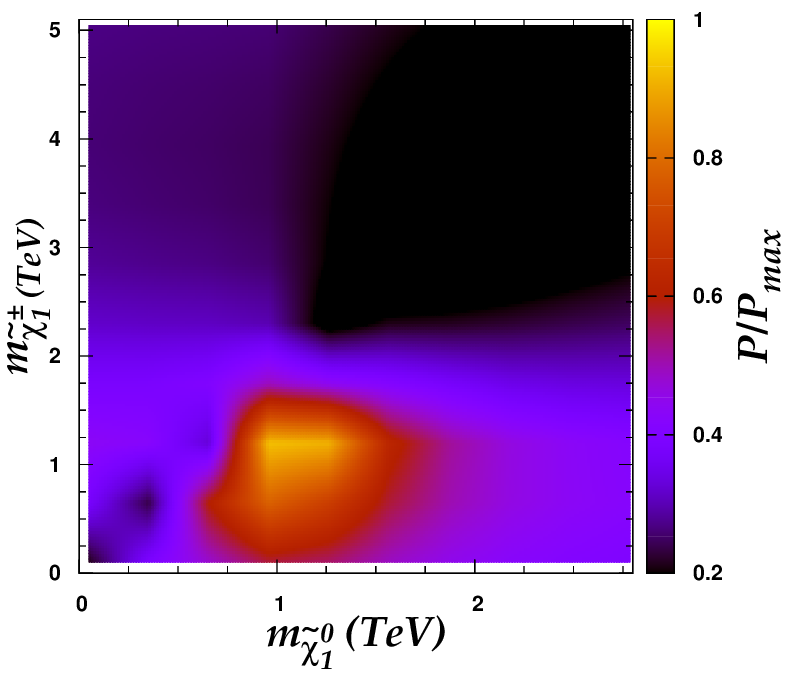}
\caption{Heatmaps of sparticles masses with posterior probability for three different sets of constraints i.e. LEP$+$HO (first row), LEP$+$HO$+$PO (second row), and LEP$+$HO$+$PO$+$DM (third row) constraints in the NMFV framework using flat priors.}
\label{fig:6}
\end{centering} 
\end{figure}

\begin{figure}
\begin{centering}
\includegraphics[angle=0,width=0.32\linewidth,height=10em]{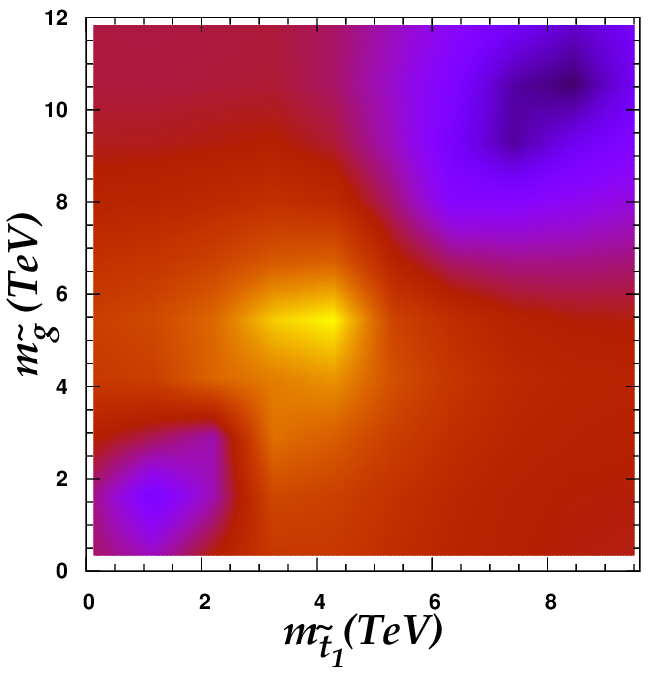}
\hspace{-0.5em}
\includegraphics[angle=0,width=0.32\linewidth,height=10em]{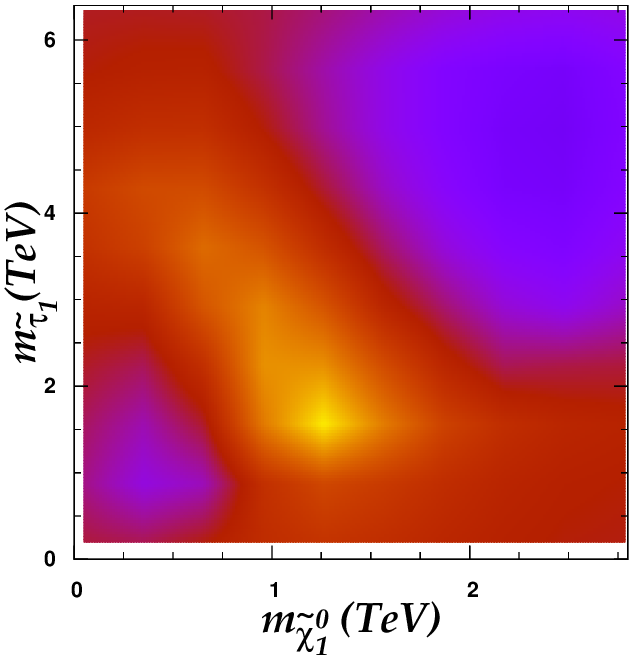}
\hspace{-0.5em}
\includegraphics[angle=0,width=0.33\linewidth,height=10em]{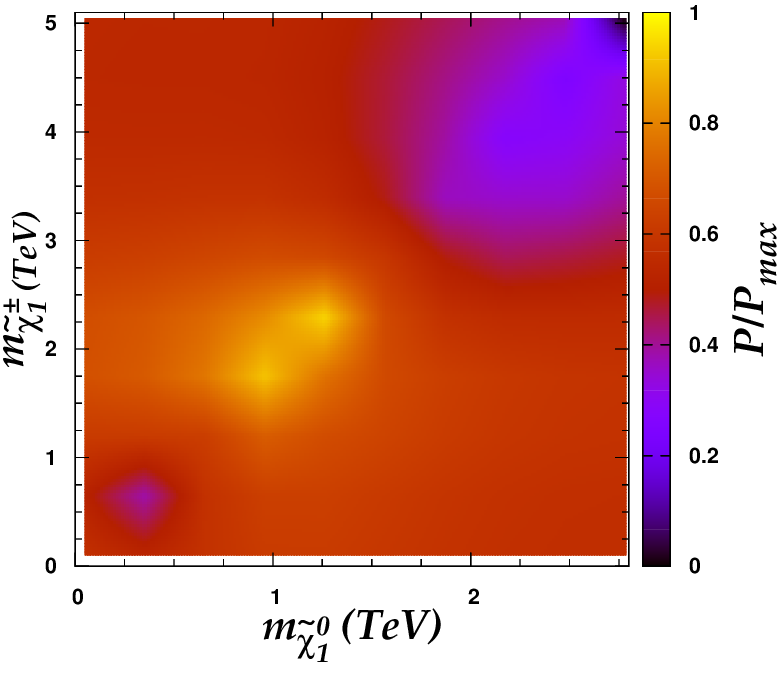}
\includegraphics[angle=0,width=0.32\linewidth,height=10em]{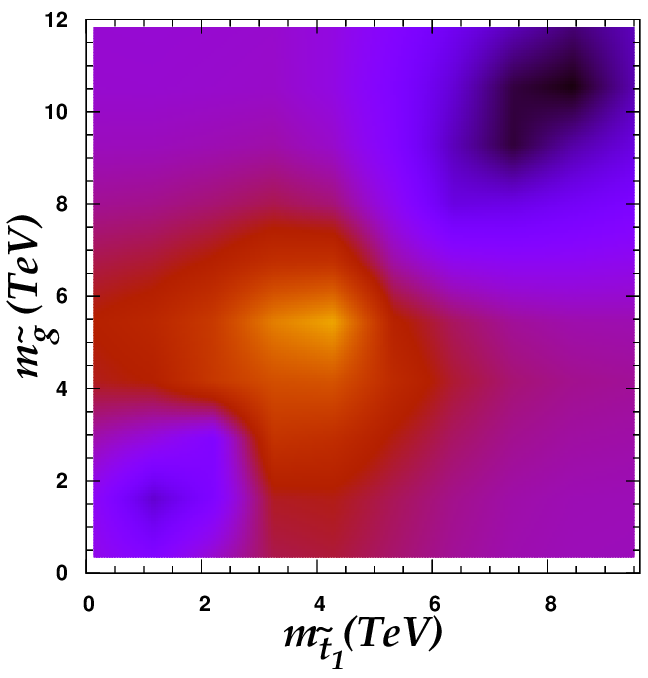}
\hspace{-0.5em}
\includegraphics[angle=0,width=0.32\linewidth,height=10em]{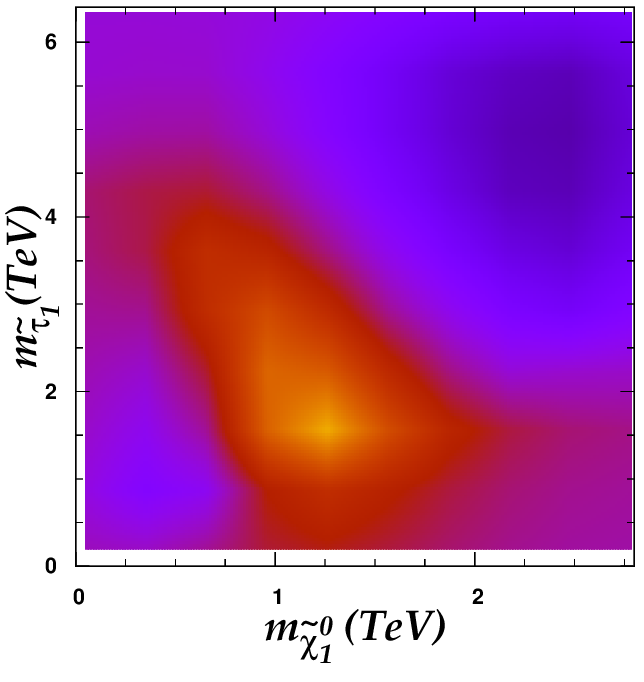}
\hspace{-0.5em}
\includegraphics[angle=0,width=0.33\linewidth,height=10em]{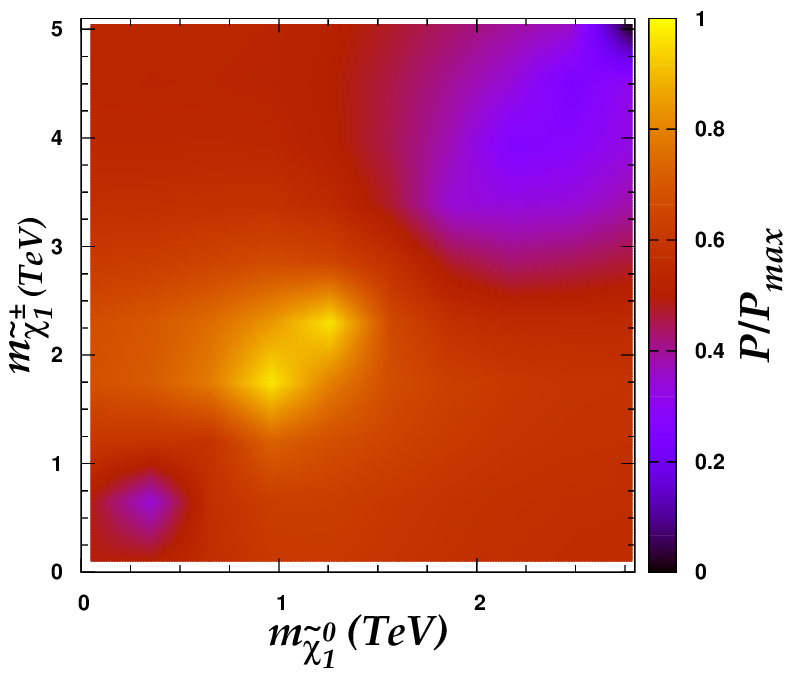}
\includegraphics[angle=0,width=0.32\linewidth,height=10em]{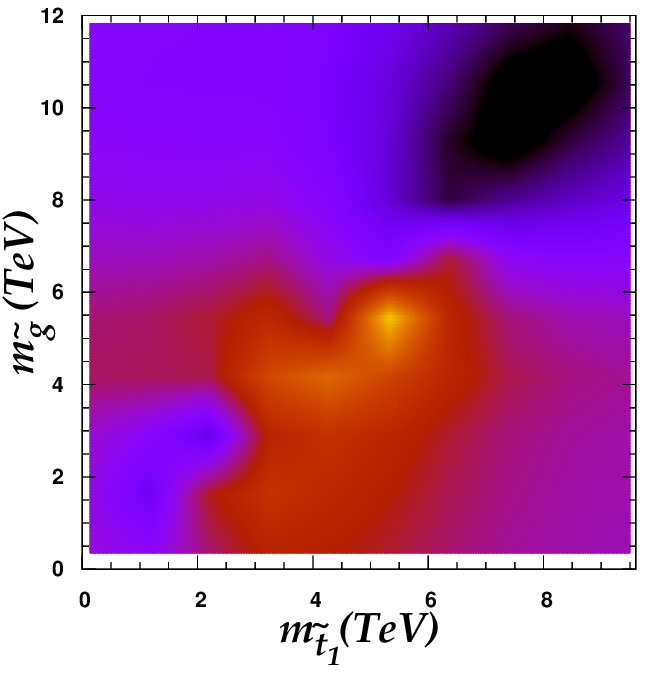}
\hspace{-0.5em}
\includegraphics[angle=0,width=0.32\linewidth,height=10em]{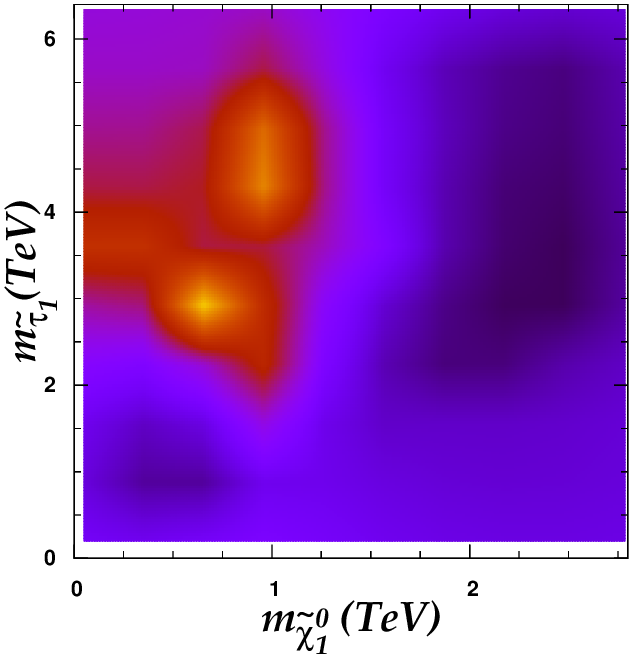}
\hspace{-0.5em}
\includegraphics[angle=0,width=0.33\linewidth,height=10em]{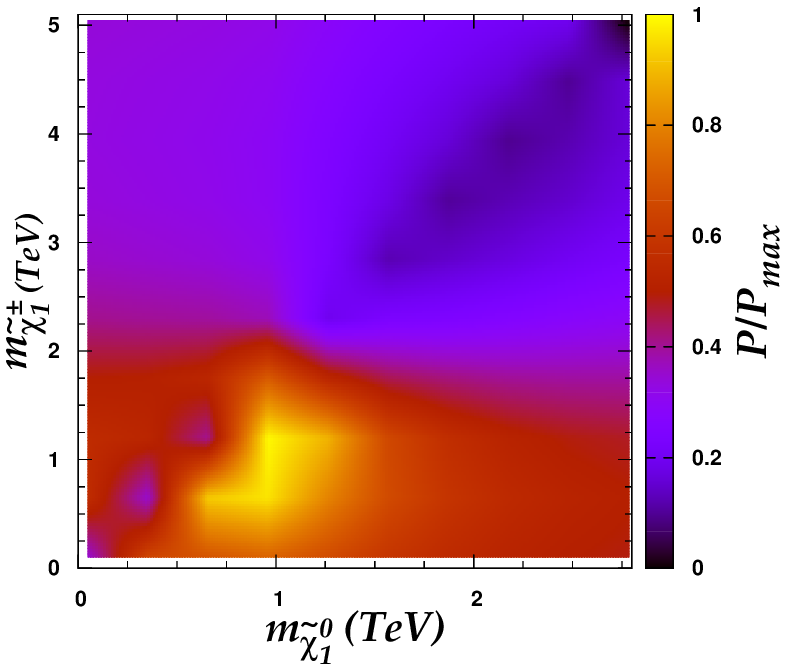}

\caption{Heatmaps of sparticles masses with posterior probability for three different sets of constraints in the NMFV framework using natural priors. The convention follows the same as in Figure~\ref{fig:6}.}
\label{fig:7}
\end{centering} 
\end{figure}

\noindent
The variation of the CMSSM parameters against the posterior probability 
for flat priors, i.e., $\xi \propto$ 1 and natural priors, $\xi 
\propto$ $(m_0 m_{1/2})^{-1}$, respectively, taking into consideration 
the constraints of LEP data, Higgs observables, precision observables, 
and the relic density of the neutralino dark matter in the $\emph{LR}$ 
sector are presented in Figures~\ref{fig:2} and \ref{fig:3}. Taking note 
of the parameter, $\tan{\beta}$, the CMSSM case reveals a high 
preference in the value of $\tan{\beta} \approx 45$ with flat priors and 
$\approx 46$ with natural priors which is expected as a higher 
$\tan{\beta}$ term contributes positively towards the Higgs mass. 
Comparatively, the reduced value of favoured regions in the NMFV case 
for $\tan{\beta}\approx 42$ with flat priors and $\approx 45$ with 
natural priors can be attributed to the phenomenon of NMFV supplementing 
the mass of the Higgs that allows for the likelihood of other regions to 
increase relatively. By comparing the two choices of models, it is 
apparent that both unified scalar and gaugino masses shift favourably 
towards the lower energy range for the NMFV case while both masses can 
be seen as naturally increasing. In the base CMSSM model, the expected 
values of $m_{0}$, $m_{1/2}$, and $A_{0}$ are 5.98 TeV, 3.37 TeV, and 
$-$0.35 TeV, respectively, with flat priors while these values correspond 
to 6.00 TeV, 3.21 TeV, and $-$0.35 TeV, respectively, with natural 
priors. In NMFV the corresponding values of $m_{0}$, $m_{1/2}$ and $A_0$ 
should be 4.83 TeV, 2.54 TeV, and 1.90 TeV, respectively, with flat 
priors and 3.25 TeV, 2.13 TeV, and 1.90 TeV, respectively, with natural 
priors. For the sake of deep understanding, 
Figures~\ref{fig:1},~\ref{fig:4}--\ref{fig:7} are presented with three 
different sets of constraints, i.e., (i) LEP$+$HO, (ii) LEP$+$HO$+$PO, 
and (iii) LEP$+$HO$+$PO$+$DM. It is to be noted that the likelihood of 
satisfying the observables peaks at $\delta^{LR}_{ct} \approx 6\times10^{-2}$ for 
flat as well as natural priors when all the constraints are included as 
shown in Figure~\ref{fig:1}. The most probable values of 
$\delta^{LR}_{ct}$ associated with different sets of constraints using 
flat and natural priors are presented in Table~\ref{tab:table4}. These 
non-zero preferable peaks motivate the belief that the NMFV supplemented 
CMSSM is able to accommodate the addition of new physics under the 
constraints of experimental observables. However, we explore our study 
only with the $\emph{LR}$ sector of the scharm-stop flavour-violating 
interaction as giving the largest contribution to the Higgs mass 
relative to other flavour-violating couplings such as 
$\delta^{RL}_{ct}$, $\delta^{LL}_{23}$, and 
$\delta^{RR}_{ct}$~\cite{AranaCatania:2011ak,AranaCatania:2012sn,Arana-Catania:2014ooa}. 
We present heatmaps of $m_{0}$, 
$m_{1/2}$ and posterior probability bounded by the different sets of 
constraints in Figures~\ref{fig:4} and \ref{fig:5} with flat and natural 
priors, respectively. Likewise, we also show heatmaps of sparticles mass 
with a posterior probability corresponding to flat and natural priors in 
Figures~\ref{fig:6} and \ref{fig:7}, respectively. The most probable 
masses of gluino, lighter stop, lighter stau, lighter chargino, and the 
lightest neutralino with flat priors taking into account all the 
above-mentioned constraints are 5.57 TeV, 5.14 TeV, 4.62 TeV, 1.07 TeV, 
and 1.04 TeV, respectively while these values are changed in natural 
priors by 4.54 TeV, 3.55 TeV, 2.47 TeV, 1.04 TeV, and 0.95 TeV, 
respectively. The most preferred values of various parameters in our 
work are listed in Table~\ref{tab:table4}. It is to be noted that the 
other way to look at the most probable masses of $m_{0}$, $m_{1/2}$ and 
sparticles can be found relatively well through heatmaps as indicated in 
Figures~\ref{fig:4}--\ref{fig:7}. The Bayes factors are found to be 5.73 
with the flat priors and 6.01 with the natural priors after taking into 
account all the above-mentioned constraints, as illustrated in 
Table~\ref{tab:table3}, predicting that the NMFV framework is strongly 
favoured against the CMSSM framework.

\vspace{-2em}

\begin{table}
\begin{centering}
	\begin{tabular}{ |l| l | l |l|}											\hline
			\multicolumn{3}{|l|}{{$ Q_{Bayes} = \log_{10}\left(\frac{{\mathcal {E}}_{NMFV}}{{\mathcal {E}}_{CMSSM}}\right)$}} 	\\
	\cline{1-4}
		{LEP$+$HO}	 	& {$+$PO} 	& {$+$DM}		\\\hline\hline
					0.02	&0.10   &5.73		\\
	0.10		&0.16		&6.01	\\
	\hline
\end{tabular}
\caption{Bayes factors in the NMFV scenario taking into account the experimental constraints with flat priors (first row) and natural priors (second row) for $m_0 \in \left[0.1, 4\right]$, $m_{1/2}
	\in \left[0.1, 4\right]$, $A_0 \in \left[-4, 4\right]$ (all in TeV units), $\tan\beta \in \left[0, 60\right]$, and $\delta^{LR}_{ct} \in \left[-0.07, 0.07\right]$.}
	\label{tab:table3}
	\end{centering}
\end{table}

\begin{table}[h!]
  \begin{center}
    \small
    \begin{tabular}{lccc|cccc} 
      \hline
      \hline
      \multirow{2}{*}{Parameter}& \multicolumn{3}{c}{Flat priors}&          \multicolumn{3}{c}{Natural priors}&\\
      \cline{2-7} 
       &{LEP$+$HO} & {$+$PO} & {$+$DM}& 
        {LEP$+$HO} & {$+$PO} & {$+$DM}\\
       \hline
      $ m_{0}$  &4.06&3.63&4.83&1.48&1.59&3.25\\
      $ m_{1/2}$  &2.42&2.36&2.54&2.39&2.34&2.13\\
      $ A_{0}$  &$-5.18$&$-4.68$&1.90&$-5.28$&$-5.00$&1.90 \\
      $ tan\beta$  &10.7&40.3&41.5&26.4&36.7&44.7\\
      $\delta^{LR}_{ct}$ &4.4$\times10^{-2}$&5$\times10^{-2}$&6.1$\times10^{-2}$&5.5$\times10^{-2}$&5.5$\times10^{-2}$&5.9$\times10^{-2}$\\
       \hline
      $m_{H}$  &3.33&3.33&2.05&3.33&3.14&2.06\\
      $m_{A^0}$  &3.52&3.33&2.04&3.33&3.23&2.04\\
      $m_{H^{\pm}}$ &3.54&3.24&2.06&3.24&3.24&2.06\\
      $ m_{\tilde\chi^{0}_{1}}$ &1.02&1.05&1.04&1.09&1.05&0.95\\
      $ m_{\tilde\chi^{0}_{2}}$  &1.96&1.96&1.09&1.96&1.96&1.09\\
       $ m_{\tilde\chi^{0}_{3}}$   &2.60&2.60&1.69&2.60&2.60&1.46\\
      $ m_{\tilde\chi^{0}_{4}}$   &2.51&2.51&1.93&2.51&2.51&1.93\\
      $ m_{\tilde\chi^{\pm}_{1}}$   &1.96&1.96&1.07&2.00&2.00&1.04\\
      $ m_{\tilde\chi^{\pm}_{2}}$  &2.62&2.52&1.93&2.52&2.52&1.93\\
      $ m_{\tilde{g}}$  &4.94&4.94 &5.57 &4.94&4.94&4.54\\
      $ m_{\tilde{q}_L}$  &5.06&4.95&7.13&4.84&4.84&4.63\\
      $ m_{\tilde{q}_R}$  &4.85&4.85&7.00&4.75&4.75&4.59\\
      $ m_{\tilde{b}_1}$   &4.34&4.34&5.95&4.34&4.34&3.98\\
      $ m_{\tilde{b}_2}$  &4.46&4.46&6.66&4.46&4.46&4.08\\
      $ m_{\tilde{t}_1}$  &3.88&3.88&5.14&3.74&3.79&3.55\\
      $ m_{\tilde{t}_2}$  &4.36&4.36&5.95&4.36&4.36&3.89\\
      $ m_{\tilde{l}_L}$  &3.34&3.07&5.71&2.63&2.63&3.46\\
      $ m_{\tilde{l}_R}$ &3.92&2.85&5.64&1.92&1.92&3.32\\
      $ m_{\tilde{\tau}_1}$ &3.79&1.95&4.62&1.49&1.49&2.47\\
       $ m_{\tilde{\tau}_2}$ &3.01&2.68&4.76&2.44&2.44&2.91\\
      \hline
      \hline
    \end{tabular}
    \caption{The sparticle mass spectrum at maximum posterior probability for NMFV framework in the $\emph{LR}$ sector of the scharm-stop flavour-violating interaction with flat and natural priors respectively. All mass parameters are in TeV except $ tan\beta$ and $\delta^{LR}_{ct}$ which are dimensionless.}
    \label{tab:table4}
 \end{center}
\end{table}

\section{Summary}
In this study, we have examined the effects of flavour-violating 
t-c interactions in the Constrained MSSM with the aid of Bayesian 
statistics. A detailed random scan over the CMSSM parameter space 
has been performed to explore the impact of such flavour-violating 
couplings by taking into account various experimental constraints 
arising from the LEP data, LHC data on Higgs, B-physics and 
electroweak precision observables, and the relic density of the 
dark matter. We have displayed the findings in 
Figures~\ref{fig:1}--\ref{fig:7}. Our work reveals that masses of
the neutralino LSP, the lighter chargino, and gluino for the 
$\emph{LR}$ sector are found to be around 1.04 TeV, 1.07 TeV, 
and 5.57 TeV, respectively, with flat priors, whereas the 
corresponding masses are observed to be about 0.95 TeV, 1.04 TeV, 
and 4.54 TeV, respectively, with natural priors. Further, sfermion 
masses are observed to be in the range of 4.62 TeV to 7.13 TeV with 
the flat priors while these masses change with the natural priors in 
the range of 2.47 TeV to 4.63 TeV. The masses of other Higgses are 
found to be around 2 TeV corresponding to flat as well as natural 
priors. Exclusively this scenario assists to reduce the SUSY breaking 
scale, thus the most probable values of free CMSSM parameters $m_{0}$, 
$m_{1/2}$, $A_{0}$, $\tan{\beta}$, and NMFV coupling parameter 
$\delta^{LR}_{ct}$ are observed to be around 4.83 TeV, 2.54 TeV, 
1.90 TeV, 41.5, and 6.1$\times10^{-2}$, respectively, with flat priors 
while these values turn out to be 3.25 TeV, 2.13 TeV, 1.90 TeV, 44.7, 
and 5.9$\times10^{-2}$, respectively, with natural priors at maximum 
posterior probability. The preferable values of $\delta^{LR}_{ct}$ 
are within top phenom constraints. The LHC Higgs mass constraint has 
favoured a small value to the flavour-violating coupling. The most 
probable values of sparticles masses, CMSSM input parameters and the 
NMFV coupling parameter in the allowed parameter space are displayed 
in Table~\ref{tab:table4}. Other NMFV effects are noteworthy in the 
case of SUSY particles, where the NMFV scenario favours lighter masses 
for SUSY particles than the base model CMSSM. The Bayes factors as 
presented in Table~\ref{tab:table3} for our model turn out to be 
about 5.73 with flat priors and 6.01 with natural priors on the 
logarithmic Jeffreys scale. While CMSSM has been taken to be the 
base model which obviously corresponds to ``decisive'' evidence on 
Jeffreys scale. Our work reveals that the most probable values of 
the neutralino LSP, corresponding to the flat and natural priors,
are found to be around 1.04 TeV and 0.95 TeV, respectively. It would be 
preferable to consider this effect at the LHC to meet the signatures of 
SUSY in the future.

\section*{Acknowledgments}
We thank Csaba Balazs for critical reading, discussions and valuable 
comments throughout the span of this work. This work was supported in 
part by University Grants Commission (UGC) under a Start-Up 
Grant no. F30-377/2017 (BSR). We acknowledge the use of computing 
facility at the DST computational lab of the Physics Department, AMU, 
Aligarh during the early phase of the work.


\begin{thebibliography}{54}

\bibitem{Djouadi:2005gi}
A.~Djouadi,
Phys.\ Rept.\ {\bf 457}, 1-216 (2008), 
[arXiv:hep-ph/0503172 [hep-ph]].


\bibitem{Aad:2015zhl}
G.~Aad {\it et al.} [ATLAS and CMS],
Phys.\ Rev.\ Lett.\ {\bf 114}, 191803 (2015),
[arXiv:1503.07589 [hep-ex]].


\bibitem{Kane:1993td}
G.~L.~Kane, C.~F.~Kolda, L.~Roszkowski and J.~D.~Wells,
Phys.\ Rev.\ D {\bf 49}, 6173-6210 (1994),
[arXiv:hep-ph/9312272 [hep-ph]].


\bibitem{allenach}
B.~C.~Allanach, K.~Cranmer, C.~G.~Lester and A.~M.~Weber,
JHEP {\bf 08}, 023 (2007),
[arXiv:0705.0487 [hep-ph]].


\bibitem{Balazs:2013qva}
C.~Balazs, A.~Buckley, D.~Carter, B.~Farmer and M.~White,
Eur.\ Phys.\ J.\ C {\bf 73}, 2563 (2013),
[arXiv:1205.1568 [hep-ph]].


\bibitem{Fowlie:2012im}
A.~Fowlie, M.~Kazana, K.~Kowalska, S.~Munir, L.~Roszkowski, E.~M.~Sessolo, S.~Trojanowski and Y.~L.~S.~Tsai,
Phys.\ Rev.\ D {\bf 86}, 075010 (2012),
[arXiv:1206.0264 [hep-ph]].


\bibitem{Athron:2017fxj}
P.~Athron, C.~Balazs, B.~Farmer, A.~Fowlie, D.~Harries and D.~Kim,
JHEP {\bf 10}, 160 (2017),
[arXiv:1709.07895 [hep-ph]].


\bibitem{GAMBIT:2017snp}
P.~Athron {\it et al.} [GAMBIT],
Eur.\ Phys.\ J.\ C {\bf 77}, no.12, 824 (2017),
[arXiv:1705.07935 [hep-ph]].


\bibitem{Ellis:2018jyl}
J.~Ellis, J.~L.~Evans, F.~Luo, K.~A.~Olive and J.~Zheng,
Eur.\ Phys.\ J.\ C {\bf 78}, no.5, 425 (2018),
[arXiv:1801.09855 [hep-ph]].


\bibitem{Drees:1996ca}
M.~Drees,
[arXiv:hep-ph/9611409 [hep-ph]].


\bibitem{Martin:1997ns}
S.~P.~Martin,
Adv.\ Ser.\ Direct.\ High Energy Phys.\ {\bf 21}, 1-153 (2010),
[arXiv:hep-ph/9709356 [hep-ph]].


\bibitem{Tata:1997uf}
X.~Tata,
[arXiv:hep-ph/9706307 [hep-ph]].


\bibitem{Chung:2003fi}
D.~J.~H.~Chung, L.~L.~Everett, G.~L.~Kane, S.~F.~King, J.~D.~Lykken and L.~T.~Wang,
Phys.\ Rept.\ {\bf 407}, 1-203 (2005),
[arXiv:hep-ph/0312378 [hep-ph]].


\bibitem{Aitchison:2005cf}
I.~J.~R.~Aitchison,
[arXiv:hep-ph/0505105 [hep-ph]].


\bibitem{Djouadi:2005} 
A.~Djouadi,
Phys.\ Rept.\ {\bf 459}, 1-241 (2008),
[arXiv:hep-ph/0503173 [hep-ph]].


\bibitem{Fayet:2015sra}
P.~Fayet,
Adv.\ Ser.\ Direct.\ High Energy Phys.\ {\bf 26}, 397-454 (2016),
[arXiv:1506.08277 [hep-ph]].


\bibitem{Allanchach:2019wrx}
B.~C.~Allanchach,
CERN Yellow Rep.\ School Proc.\ {\bf 6}, 113-144 (2019).


\bibitem{Cane:2019ac}
A.~Canepa,
Rev.\ Phys.\ {\bf 4}, 100033 (2019).


\bibitem{Balazs:2012bx}
C.~Bal\'azs and S.~K.~Gupta,
Phys.\ Rev.\ D {\bf 87}, no.3, 035023 (2013),
[arXiv:1212.1708 [hep-ph]].


\bibitem{Heinemeyer:2004by}
S.~Heinemeyer, W.~Hollik, F.~Merz and S.~Penaranda,
Eur.\ Phys.\ J.\ C {\bf 37}, 481-493 (2004),
[arXiv:hep-ph/0403228 [hep-ph]].


\bibitem{Bozzi:2007me}
G.~Bozzi, B.~Fuks, B.~Herrmann and M.~Klasen,
Nucl.\ Phys.\ B {\bf 787}, 1-54 (2007),
[arXiv:0704.1826 [hep-ph]].


\bibitem{AranaCatania:2011ak}
M.~Arana-Catania, S.~Heinemeyer, M.~J.~Herrero and S.~Penaranda,
JHEP {\bf 05}, 015 (2012),
[arXiv:1109.6232 [hep-ph]].


\bibitem{AranaCatania:2012sn}
M.~Arana-Catania, S.~Heinemeyer, M.~J.~Herrero and S.~Penaranda,
[arXiv:1201.6345 [hep-ph]].


\bibitem{Arana-Catania:2014ooa}
M.~Arana-Catania, S.~Heinemeyer and M.~J.~Herrero,
Phys.\ Rev.\ D {\bf 90}, no.7, 075003 (2014),
[arXiv:1405.6960 [hep-ph]].


\bibitem{Bernigaud:2018qky}
J.~Bernigaud, B.~Herrmann, S.~F.~King and S.~J.~Rowley,
JHEP {\bf 03}, 067 (2019),
[arXiv:1812.07463 [hep-ph]].


\bibitem{Kowalska:2014opa}
K.~Kowalska,
JHEP {\bf 09}, 139 (2014),
[arXiv:1406.0710 [hep-ph]].


\bibitem{Bernigaud:2018vmh}
J.~Bernigaud and B.~Herrmann,
SciPost Phys.\ {\bf 6}, no.6, 066 (2019),
[arXiv:1809.04370 [hep-ph]].


\bibitem{DeCausmaecker:2015yca}
K.~De Causmaecker, B.~Fuks, B.~Herrmann, F.~Mahmoudi, B.~O'Leary, W.~Porod, S.~Sekmen and N.~Strobbe,
JHEP {\bf 11}, 125 (2015),
[arXiv:1509.05414 [hep-ph]].


\bibitem{Carena:2006ai}
M.~Carena, A.~Menon, R.~Noriega-Papaqui, A.~Szynkman and C.~E.~M.~Wagner,
Phys.\ Rev.\ D {\bf 74}, 015009 (2006),
[arXiv:hep-ph/0603106 [hep-ph]].


\bibitem{delAguila:2008iz}
F.~del Aguila, J.~A.~Aguilar-Saavedra, B.~C.~Allanach, J.~Alwall, Y.~Andreev, D.~Aristizabal Sierra, A.~Bartl, M.~Beccaria, S.~Bejar and L.~Benucci, \textit{et al.},
Eur.\ Phys.\ J.\ C {\bf 57}, 183-308 (2008).
[arXiv:0801.1800 [hep-ph]].


\bibitem{Bruhnke:2010rh}
M.~Bruhnke, B.~Herrmann and W.~Porod,
JHEP {\bf 09}, 006 (2010),
[arXiv:1007.2100 [hep-ph]].


\bibitem{Herrmann:2011xe}
B.~Herrmann, M.~Klasen and Q.~Le Boulc'h,
Phys.\ Rev.\ D {\bf 84}, 095007 (2011),
[arXiv:1106.6229 [hep-ph]].


\bibitem{Hahn:2005qi}
T.~Hahn, W.~Hollik, J.~I.~Illana and S.~Penaranda,
[arXiv:hep-ph/0512315 [hep-ph]].


\bibitem{Fuks:2008ab}
B.~Fuks, B.~Herrmann and M.~Klasen,
Nucl.\ Phys.\ B {\bf 810}, 266-299 (2009),
[arXiv:0808.1104 [hep-ph]].


\bibitem{Fuks:2011dg}
B.~Fuks, B.~Herrmann and M.~Klasen,
Phys.\ Rev.\ D {\bf 86}, 015002 (2012),
[arXiv:1112.4838 [hep-ph]].


\bibitem{Hu:2019heu}
Q.~Y.~Hu, X.~Q.~Li, Y.~D.~Yang and M.~D.~Zheng,
JHEP {\bf 06}, 133 (2019),
[arXiv:1903.06927 [hep-ph]].


\bibitem{Gupta:2022bjx}
S.~Gupta and S.~K.~Gupta,
Nucl.\ Phys.\ B {\bf 984}, 115942 (2022),
[arXiv:2205.00173 [hep-ph]].


\bibitem{Hiller:2008sv}
G.~Hiller, Y.~Hochberg and Y.~Nir,
JHEP {\bf 03}, 115 (2009),
[arXiv:0812.0511 [hep-ph]].


\bibitem{Guasch:1999jp}
J.~Guasch and J.~Sola,
Nucl.\ Phys.\ B {\bf 562}, 3-28 (1999),
[arXiv:hep-ph/9906268 [hep-ph]].


\bibitem{Cao:2006xb}
J.~Cao, G.~Eilam, K.~i.~Hikasa and J.~M.~Yang,
Phys.\ Rev.\ D {\bf 74}, 031701 (2006),
[arXiv:hep-ph/0604163 [hep-ph]].


\bibitem{Cao:2007dk}
J.~J.~Cao, G.~Eilam, M.~Frank, K.~Hikasa, G.~L.~Liu, I.~Turan and J.~M.~Yang,
Phys.\ Rev.\ D {\bf 75}, 075021 (2007),
[arXiv:hep-ph/0702264 [hep-ph]].


\bibitem{Curiel:2002pf}
A.~M.~Curiel, M.~J.~Herrero and D.~Temes,
Phys.\ Rev.\ D {\bf 67}, 075008 (2003),
[arXiv:hep-ph/0210335 [hep-ph]].


\bibitem{Curiel:2003uk}
A.~M.~Curiel, M.~J.~Herrero, W.~Hollik, F.~Merz and S.~Penaranda,
Phys.\ Rev.\ D {\bf 69}, 075009 (2004),
[arXiv:hep-ph/0312135 [hep-ph]].


\bibitem{Bejar:2004rz}
S.~Bejar, F.~Dilme, J.~Guasch and J.~Sola,
JHEP {\bf 08}, 018 (2004),
[arXiv:hep-ph/0402188 [hep-ph]].


\bibitem{Atwood:2013ica}
D.~Atwood, S.~K.~Gupta and A.~Soni,
JHEP {\bf 10}, 057 (2014),
[arXiv:1305.2427 [hep-ph]].


\bibitem{Zyla:2020zbs}
P.~A.~Zyla {\it et al.} [Particle Data Group], 
Prog.\ Theor.\ Exp.\ Phys.\ {\bf 2020}, no.8, 083C01 (2020).


\bibitem{Allanach:2001kg}
B.~C.~Allanach,
Comput.\ Phys.\ Commun.\ {\bf 143}, 305-331 (2002),
[arXiv:hep-ph/0104145 [hep-ph]].


\bibitem{Heinemeyer:1998np}
S.~Heinemeyer, W.~Hollik and G.~Weiglein,
Eur.\ Phys.\ J.\ C {\bf 9}, 343-366 (1999),
[arXiv:hep-ph/9812472 [hep-ph]].


\bibitem{Heinemeyer:1998yj}
S.~Heinemeyer, W.~Hollik and G.~Weiglein,
Comput.\ Phys.\ Commun.\ {\bf 124}, 76-89 (2000),
[arXiv:hep-ph/9812320 [hep-ph]].


\bibitem{Degrassi:2002fi}
G.~Degrassi, S.~Heinemeyer, W.~Hollik, P.~Slavich and G.~Weiglein,
Eur.\ Phys.\ J.\ C {\bf 28}, 133-143 (2003),
[arXiv:hep-ph/0212020 [hep-ph]].


\bibitem{Frank:2006yh}
M.~Frank, T.~Hahn, S.~Heinemeyer, W.~Hollik, H.~Rzehak and G.~Weiglein,
JHEP {\bf 02}, 047 (2007),
[arXiv:hep-ph/0611326 [hep-ph]].


\bibitem{Hahn:2013ria}
T.~Hahn, S.~Heinemeyer, W.~Hollik, H.~Rzehak and G.~Weiglein,
Phys.\ Rev.\ Lett.\ {\bf 112}, no.14, 141801 (2014),
[arXiv:1312.4937 [hep-ph]].


\bibitem{Bahl:2016brp}
H.~Bahl and W.~Hollik,
Eur.\ Phys.\ J.\ C {\bf 76}, no.9, 499 (2016),
[arXiv:1608.01880 [hep-ph]].


\bibitem{Bahl:2017aev}
H.~Bahl, S.~Heinemeyer, W.~Hollik and G.~Weiglein,
Eur.\ Phys.\ J.\ C {\bf 78}, no.1, 57 (2018),
[arXiv:1706.00346 [hep-ph]].


\bibitem{Bahl:2018qog}
H.~Bahl, T.~Hahn, S.~Heinemeyer, W.~Hollik, S.~Paßehr, H.~Rzehak and G.~Weiglein,
Comput.\ Phys.\ Commun.\ {\bf 249}, 107099 (2020),
[arXiv:1811.09073 [hep-ph]].



\bibitem{Arbey:2018msw}
A.~Arbey, F.~Mahmoudi and G.~Robbins,
Comput.\ Phys.\ Commun.\ {\bf 239}, 238-264 (2019),
[arXiv:1806.11489 [hep-ph]].


\bibitem{Belanger:2004yn}
G.~Belanger, F.~Boudjema, A.~Pukhov and A.~Semenov,
Comput.\ Phys.\ Commun.\ {\bf 174}, 577-604 (2006),
[arXiv:hep-ph/0405253 [hep-ph]].


\bibitem{Belanger:2001fz}
G.~Belanger, F.~Boudjema, A.~Pukhov and A.~Semenov,
Comput\. Phys.\ Commun.\ {\bf 149}, 103-120 (2002),
[arXiv:hep-ph/0112278 [hep-ph]].


\bibitem{Schael:2006cr}
S.~Schael {\it et al.} [ALEPH, DELPHI, L3, OPAL and LEP Working Group for Higgs Boson Searches],
Eur.\ Phys.\ J.\ C {\bf 47}, 547-587 (2006),
[arXiv:hep-ex/0602042 [hep-ex]].



\bibitem{Abi:2021gix}
B.~Abi {\it et al.} [Muon g-2],
Phys.\ Rev.\ Lett.\ {\bf 126}, no.14, 141801 (2021),
[arXiv:2104.03281 [hep-ex]].


\bibitem{Amhis:2019ckw}
Y.~S.~Amhis {\it et al.} [HFLAV],
Eur.\ Phys.\ J.\ C {\bf 81}, no.3, 226 (2021),
[arXiv:1909.12524 [hep-ex]].


\bibitem{Planck:2015fie}
P.~A.~R.~Ade \textit{et al.} [Planck],
Astron.\ Astrophys.\ {\bf 594}, A13 (2016),
[arXiv:1502.01589 [astro-ph.CO]].


\end{thebibliography}
\end{document}